\documentclass[conference]{IEEEtran}
\IEEEoverridecommandlockouts
\usepackage{cite}
\usepackage{amsmath,amssymb,amsfonts}
\usepackage{algorithm}
\usepackage{algpseudocode}
\usepackage{graphicx}
\usepackage{textcomp}
\usepackage{xcolor}
\usepackage{algpseudocode}
\usepackage{subfigure}
\def\BibTeX{{\rm B\kern-.05em{\sc i\kern-.025em b}\kern-.08em
    T\kern-.1667em\lower.7ex\hbox{E}\kern-.125emX}}

\newboolean{showcomments}
\setboolean{showcomments}{true}
\ifthenelse{\boolean{showcomments}}
{
}

\newcommand*{\affmark}[1][*]{\textsuperscript{#1}}

\begin{document}

\title{Traffic Steering for 5G Multi-RAT Deployments using Deep Reinforcement Learning }

\author{\IEEEauthorblockN{Md~Arafat~Habib\affmark[1], Hao~Zhou\affmark[1], Pedro~Enrique~Iturria-Rivera\affmark[1], Medhat Elsayed\affmark[2], Majid Bavand\affmark[2], \\Raimundas Gaigalas\affmark[2], Steve Furr\affmark[2] and  Melike Erol-Kantarci\affmark[1], \IEEEmembership{Senior Member,~IEEE}}
\IEEEauthorblockA{\affmark[1]\textit{School of Electrical Engineering and Computer Science, University of Ottawa, Ottawa, Canada}}  \affmark[2]\textit{Ericsson Inc., Ottawa, Canada}\\
Emails:\{mhabi050, hzhou098, pitur008, melike.erolkantarci\}@uottawa.ca, \\\{medhat.elsayed, majid.bavand, raimundas.gaigalas, steve.furr\}@ericsson.com \vspace{-1em}}

\maketitle

\begin{abstract}
In 5G non-standalone mode, traffic steering is a critical technique to take full advantage of 5G new radio while optimizing dual connectivity of 5G and LTE networks in multiple radio access technology (RAT). An intelligent traffic steering mechanism can play an important role to maintain seamless user experience by choosing appropriate RAT (5G or LTE) dynamically for a specific user traffic flow with certain QoS requirements. In this paper, we propose a novel traffic steering mechanism based on Deep Q-learning that can automate traffic steering decisions in a dynamic environment having multiple RATs, and maintain diverse QoS requirements for different traffic classes. The proposed method is compared with two baseline algorithms: a heuristic-based algorithm and Q-learning-based traffic steering. Compared to the Q-learning and heuristic baselines, our results show that the proposed algorithm achieves better performance in terms of 6\% and 10\% higher average system throughput, and 23\% and 33\% lower network delay, respectively.
\end{abstract}


\section{Introduction}
The dual connectivity between long term evolution (LTE) and fifth generation new radio (5G NR) results in multiple radio access technologies (multi-RAT)\cite{16,20}. On the other hand, each type of RAT is supposed to have distinctive capabilities to serve user equipment (UE) with diverse quality-of-service (QoS) requirements. This raises the need of steering a specific class of traffic to a certain RAT to fulfill the QoS demands. For instance, high throughput video traffic can be better served by 5G NR. On the contrary, steering voice traffic to LTE base station (BS) with wider coverage can be a better decision since such traffic is not throughput hungry but requires more coverage to avoid frequent handovers. However, steering a specific class of traffic continuously to a certain RAT may cause several problems. The system may suffer from higher delay due to excessive load and reduced throughput because of the packet drops. These issues are quite challenging to address, especially when 5G NR facilitates dense network deployments and an increased number of users.

To address the above-mentioned challenges, an AI-enabled traffic steering scheme emerges as a promising approach to manage densely deployed networks with dynamic requirements. In recent years, AI and machine learning have been applied to various other problems in 5G \cite{21}. Even though the emergence of the 5G non-stand-alone (NSA) mode has drawn the attention of researchers recently, most existing works linked with traffic steering lack a comprehensive tool to overcome the complexity.

For instance, in \cite{1}, the authors propose a traffic steering scheme based on some threshold calculated using parameters like load at each type of RAT, channel condition, and service type but the method lacks the intelligence to handle dynamic wireless environments. Compared with conventional model-based optimization methods, machine learning, especially reinforcement learning (RL) algorithms, can significantly reduce the complexity of defining a dedicated optimization model\cite{bz1}. Advanced machine learning techniques like deep reinforcement learning (DRL) \cite{17} can not only automate traffic steering in a dynamic 5G wireless environment, but also it can handle larger state-action space compared to traditional reinforcement learning. Therefore, unlike previous works, we propose a DRL-based traffic steering scheme that tends to perform RAT specific traffic steering in a multi-RAT environment to maintain QoS requirements of different traffic classes in a dynamic 5G NSA mode to maintain seamless network activity and smooth user experience.   

In this paper, we seek to balance the QoS demands of all the traffic classes simultaneously by proposing a Deep-Q-network (DQN)-based traffic steering scheme. The reward and state functions of the proposed DQN-based traffic steering scheme is carefully designed to have satisfactory performance based on two crucial key performance indicators (KPIs); i.e. network delay and average system throughput. Performance of the proposed method is compared with two baseline algorithms: Q-learning-based method \cite{3} and a heuristic-based algorithm adopted from \cite{1}. It gains 6\% and 10\% increase in average system throughput compared to the Q-learning and heuristic-based baseline respectively. Furthermore, it achieves 23\% and 33\% decrease in network delay compared to the mentioned baselines.  

The rest of the paper is organized as follows: Section \ref{s2} presents the related works. We discuss the system model and the problem formulation in Section \ref{s3}. Section \ref{s4} covers the proposed DQN-based traffic steering scheme along with the baselines. The performance evaluation of the proposed DQN-based traffic steering method is presented in Section \ref{s5}. Finally, the paper is concluded in Section \ref{s6}.

\section{Related works}
\label{s2}
In this section, we summarize the state-of-the-art literature on traffic steering. Prasad et al. propose a dynamic traffic steering scheme for energy efficient radio access network moderation in ultra-dense 5G networks \cite{19}. A unified traffic steering scheme by Dryjanski et al. is proposed for LTE-advanced pro, aiming at optimal radio resource allocation in multi-RAT networks \cite{18}. Most recently, Khaled et al. have proposed a cell zooming technique to steer traffic in a software defined radio-enabled LTE network that uses renewable energy sources to lessen on-grid power consumption \cite{9}. Gijon et al. propose a data driven approach to perform traffic steering in multi-carrier LTE networks in which traffic steering is conducted based on reference signal received quality-based handover margins \cite{10}.

Nevertheless, 5G deployments have made it more challenging to develop an elegant traffic steering scheme because of the increased number of users and dual connectivity. Passas et al. propose a pricing oriented network selection process for distributed heterogeneous networks based on imposed load pressure at a particular RAT \cite{11}. A heuristic-based approach proposed in \cite{1} performs traffic steering based on a threshold level calculated using parameters like channel condition, load level at each RAT, and service type. Priscoli et al. address the problem of traffic steering using a Q-learning-based solution that aims at maintaining QoS, and performs load balancing in a 5G heterogeneous network \cite{2}. Different from the previous works, this paper provides automation in the system via DRL-based traffic steering scheme that can perform RAT specific traffic steering in a multi-RAT environment. Furthermore, the proposed method can maintain QoS requirements of different traffic classes in a dynamic 5G NSA mode to maintain seamless network activity and smooth user experience.

\section{System Model and Problem Formulation}
\label{s3}
\subsection{System Model}
In this work, a multi-RAT network is considered having $Q$ classes of RATs where each class of RAT, $q$ represents a particular access technology (LTE, 5G, etc.). Multiple users are associated with different types of RATs via dual connectivity. A UE can maintain $K$ types of traffic classes. Fig. 1 presents the network model considered in this study.  We represent three different classes of traffics: voice, gaming, and video as TC1, TC2, and TC3 respectively in the figure. We have designed our network environment in a way where small cells are within the range of a macro-cell. UEs have dual connectivity with LTE or 5G RAT and traffic can be steered to either one of these RATs based on our proposed method.
\begin{figure}[htbp]
\centerline{\includegraphics[width=0.7\linewidth]{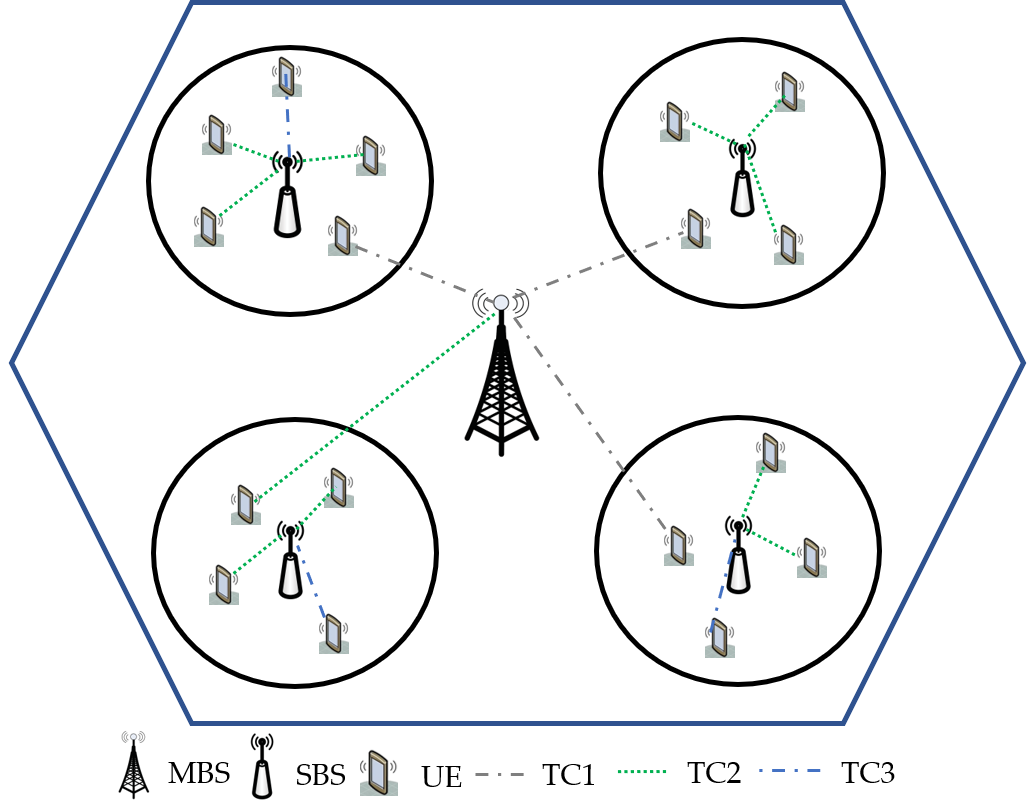}}
\caption{Illustration of network environment with one LTE macro cell and several 5G small cells.}
\label{fig}
\vspace{-1em}
\end{figure}
The total downlink bandwidth, $B$ in MHz is divided into $N_{RB}$ resource blocks. A resource block contains a set of 12 contiguous subcarriers. Consecutive resource blocks are grouped to constitute resource block group (RBG) as defined in \cite{21}. Each RBG, $h$ is allocated a certain transmission power $p_{h,b}$, by a BS, $b$. Based on our system model, each BS holds a number of transmission buffers corresponding to the number of users connected to it. Every transmission time interval (TTI), the downlink scheduler assigns resources to the users having pending data transmissions.

The link capacity between the UE, $u$ and BS, $b$ can be formulated as follows:
\begin{equation}
    C_{u,b}=\sum_{h=1}^{H}\omega_h\log_2\left(1+\frac{p_{h,b}x_{h,u,b}g_{h,u,b}}{\omega_h N_0+\sum_{m\in B}p_{h,m}x_{h,u,m}g_{h,u,m}}\right),
\end{equation}
where $\omega_h$ is the bandwidth of the $h$, $p_{h,b}$ is the transmit power of the BS, $b$ on $h$, $g_{h,u,b}$ is the channel co-efficient and $x_{h,u,b}$ is the RBG's allocation indicator of the link $(h,u,b)$. $N_0$ is the additive white Gaussian noise single-sided power spectral density. $p_{h,m}$ is the transmit power of the interfering BS, $m$, $g_{h,u,m}$ is the channel co-efficient, and $x_{h,u,m}$ is the allocation indicator of link $(h,u,m)$. 

Each link has a capacity limit. Traffic flows passing through a link should not exceed the capacity of the link in the system.
\begin{equation}
    \sum_{f\in F}d^{f}x_{u,b}^{f}\leqslant C_{u,b}\quad \forall(u,b)\in L,
\end{equation}
where $F$ is the set of all the flows in the network, $d^f$ is the capacity demand of the flow $f \in F$ from UE, $u$ to BS $b$. $x_{u,b}^{f}$ represents a binary $(0,1)$ component that is `1' if the link $(u,b)$ has been used from UE,$u$ to BS $b$. It is `0' otherwise. $L$ is the set of links and $C_{u,b}$ is the capacity of link $(u,b)$. as presented in eq. (1)

In our system model, the delay is considered as the summation of transmission and queuing delay which is as follows:
\begin{equation}
    D_{k,b}=D_{k,b}^{Trx}+D_{k,b}^q,
\end{equation}
where $D_{k,b}^{Trx}$ is the transmission delay experienced for a particular traffic type $k$ and BS $b$, and $D_{k,b}^q$ is the queuing delay experienced for a particular traffic type $k$ at BS $b$ for a user $u$. The transmission delay can be calculated as follows: 
\begin{equation}
    D_{k,b}^{Trx}=\frac{L_{u,b}}{C_{u,b}},
\end{equation}
where $L_{u,b}$ is the packet length and $C_{u,b}$ is the link capacity as stated in eq. (1).

\subsection{QoS Requirements and Problem Formulation}

To be able to perform traffic steering for different traffic classes with QoS requirements for delay and throughput, first two parameters are defined based on delay and throughput. The delay parameter associated with our traffic steering problem is considered as the ratio of the defined QoS requirement for delay and the actual delay experienced in the system for a particular traffic class being carried by a certain BS. It can be stated as follows: 
\begin{equation}
    r_{k,b}^D=\frac{D_{QoS}}{D_{k,b}},
\end{equation}
where $D_{QoS}$ is delay requirement defined in the simulation for a particular traffic type and $D_{k,b}$ is the actual delay achieved. Similarly, the throughput parameter is defined as the ratio of actual throughput achieved and the required throughput as stated in eq. (6):
\begin{equation}
    r_{k,b}^T=\frac{T_{k,b}}{T_{QoS}},
\end{equation}
where $T_{QoS}$ is the throughput requirement defined in the simulation for a particular traffic class and $T_{k,b}$ is the actual throughput achieved.

Since our aim is to improve the system performance in terms of the delay and throughput, a new variable is formed to represent and meet such targets. It combines the delay and throughput parameters in eq. (5) and (6) along with some weight factors. The declared variable combined with delay, throughput, and weight factors ($w_1$ and $w_2$) is as follows:
\begin{equation}
    M=w_1(r_{k,b}^D)+w_2(r_{k,b}^T).
\end{equation}
The traffic steering problem proposed in this paper is formulated as the maximization of the variable $M$ (presented in eq. (7)) which is as follows:
\begin{equation}
     \begin{split}
      max\sum_{u\in U}\sum_{k\in K}\sum_{b\in B}M_{u,f,b}, \quad \quad \\
     s.t. \sum_{(u,b)\in L}\beta^{f_k}\geqslant \beta^f \quad \forall f \in F, \quad \\
     \sum_{(u,b)\in L}D(u,b)x_{u,b}^f\leqslant D^f \quad \forall f \in F, \\
     \end{split}
\end{equation}
where $\beta^{f_k}$ is the required bitrate for a particular type of traffic $k$, and $\beta^f$ is the available bitrate. Also, $D^f$ represents the latency demand of flow $f\in F$ and $D(u,b)$ is the latency of link $(u,b)$.   

\section{Proposed DQN-based Traffic Steering Scheme}
\label{s4}
\subsection{DQN-based Traffic Steering Scheme}
For a relatively simplistic RL environment, Q-learning is a good solution for optimization. However, as the state-space increases, the time needed to traverse all these states and iteratively update all the Q-values will increase which is computationally inefficient and resource consuming. To address this issue, DQN can be used to estimate the Q-values for each state-action pair in a given environment using a deep neural network (DNN)\cite{17}. 

During the training stage of DQN, agent’s experiences at each time step is stored in a data set called the replay memory. At time $\tau$, the agent’s experience $e_\tau$ is defined as the following tuple:
\begin{equation}
    e_\tau=(S_\tau,A_\tau,R_{\tau+1},S_{\tau+1}).
\end{equation}
The tuple contains the state of the environment, the action taken from the state, the reward given to the agent as a result of previous state-action pair and the next state of the environment. In short, the tuple gives us the summary of the agent’s experience at time $\tau$. All the agent’s experiences at each time step over all the episodes played by the agent are stored in the replay memory. In practice, the replay memory is set to some finite size unit $(N)$. Therefore, it will only store the last $N$ experiences. The replay memory data set is the place from where random samples are chosen to train the network.

The DNN in DQN takes states as inputs from the environment and outputs the Q-values for each action that can be taken from that state. Before the training starts, first, the replay memory data set, $D$ is initialized to capacity, $N$. Next, DNN is initialized with random weights. For each episode, the starting state is initialized. For each time step within the episode, the agent either explores the environment and selects a random action or the agent exploits the environment and selects the greedy action for the given state that provides the highest Q-value. This epsilon greedy policy is used to balance the exploration and exploitation.
\begin{equation}
  A_\tau=\begin{cases}
    random \quad action, & \text{if $rand \leqslant \epsilon$} \\
    argmax(q_\tau(S_\tau,A_\tau)), & \text{otherwise}
  \end{cases}
\end{equation}
where $\epsilon$ is the exploration probability within $0\leqslant \epsilon \leqslant 1$ and $rand$ represents a random number between 0 to 1.

After an action is taken, we observe the reward for the action along with the next state of the environment. Therefore, the state an agent initialized from, action taken, reward observed are all put together in a tuple as described in eq. (9). 

For a single sample, the first pass to the network occurs for the state from the experience tuple that was sampled. The network then outputs the Q-values associated with each possible action that can be taken from that state and then the loss is calculated between the Q-values for the action from the experience tuple and the target Q-value for this action. To calculate the target Q-value, it is required to have a second pass to the target network with the next state. The target network is the clone of the policy network (which is also the main network). Its weights are frozen with the weights same as the policy network and the weights are updated in the target network after every certain amount of time steps. The loss for DQN is calculated using the following equation:
\begin{equation}
    L(w)=Er(R_\tau+\gamma \max_A q(S_{\tau+1},A,w^\prime)-q(S_\tau,A_\tau,w)),
\end{equation}
where $w$ and $w^\prime$ are the weights of the main and the target network, and $Er$ represents the error function. Having two NNs (main and target) ensures stability.

Fig. 2 describes the schematic of the proposed DQN-based traffic steering where we have a main network and a target network and minibatch from the replay memory is getting fetched.

\begin{figure}[htbp]
\centerline{\includegraphics[width=2.5in]{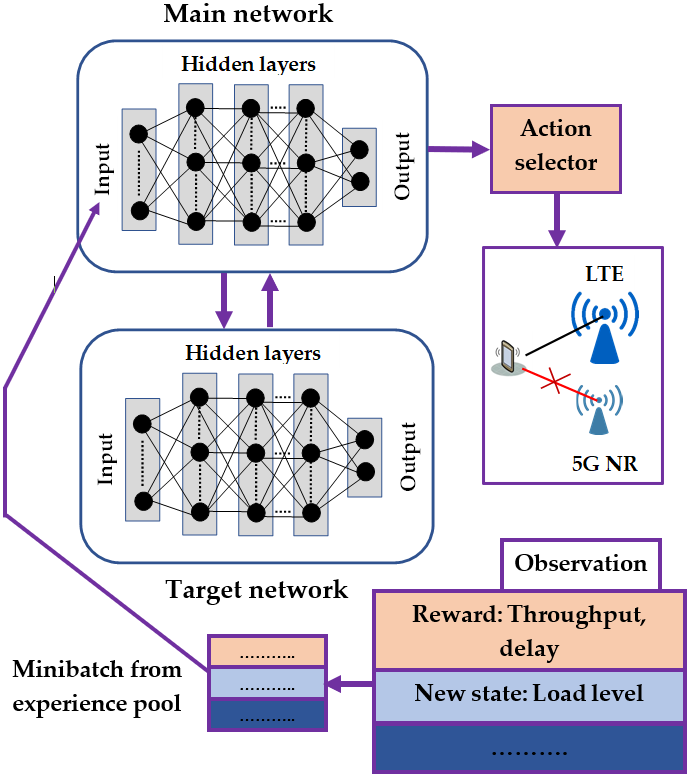}}
\caption{Overall system architecture with DQN.}
\label{fig}
\vspace{-1em}
\end{figure}

The mathematical formulation of DQN depends on Markov Decision Process (MDP) that is defined by agents, states, actions, and a reward function. Tuples associated with DQN is defined as follows:
\begin{itemize}
\item \textbf{Agent:} We implement a centralized agent to control the macro base station (MBS) and the small cell base stations. It is deployed in the MBS and controls all the incoming traffic to each BS.
\item \textbf{State:} The state consists of three elements, \{$T_f,L_{Q(SINR)},q_L$\}. Here, $T_f$ represents the traffic type. It is assumed that each traffic type has fixed QoS requirements and we can perform traffic steering to a particular RAT based on that. Users periodically report signal-to-interference and noise ratio (SINR) measurements to the 5G base station (gNB) and LTE base station (eNB). It indicates the quality of the link associated with a UE and a BS. Therefore, the second element of state space is: $L_{Q(SINR)}$=\{$SINR_{eNB},SINR_{gNB}$\}. To represent load level, queue length of both types of RATs is used. So, the last element of the state space is queue length, $q_L$=\{$q_{L(gNB)},q_{L(eNB)}\}$.
\item \textbf{Action:} The action space contains the action of flow admission to the RATs. It is defined as: \{$A_{LTE},A_{5G}\}$. Here, $(A_{LTE})$ stands for flow admission to the LTE RAT , and $(A_{5G})$ stands for flow admission to the 5G RAT.
\item \textbf{Reward:} The reward function is based on eq. (7). To keep it normalized, sigmoid function is used. Therefore, the reward function is as follows:
\begin{equation}
    R=sigm (M),
\end{equation}
where $sigm(M)$ represents the sigmoid function. 
\end{itemize}

The proposed DQN-based traffic steering algorithm is summarized as Algorithm 1.

\begin{algorithm}
\textbf {Initialize}: Network and DQN parameters
\begin{algorithmic}[1]
\caption{DQN-based traffic steering}\label{alg:cap}
\For {$TTI=1 \quad to \quad T$}
\For{every $u,b,k$}
\If{$(rand\leq \epsilon)$} 
    \State choose action randomly
\Else
    \State select $A_\tau$ using greedy policy
\EndIf
\State BSs are selected for all the UEs for all $k \in K$
\State Traffic admission is performed
\State Reward calculation based on eq. (12)
\State Agent updates its own state $S_\tau$
\State Save $(S_\tau,A_\tau,R_{\tau+1},S_{\tau+1})$
\EndFor
\State Random sample a minibatch from the experience pool
\State Generate target Q-values, $q_\tau(S_\tau,A_\tau)$
\State Update $w$ using gradient descent to minimize the loss,

$L(w)=Er(q_\tau(S_\tau,A_\tau)-q(S_\tau,A_\tau,w))$
\State Copy $w$ to $w^\prime$ after several training
\EndFor
\State \textbf {Output}: Optimal traffic steering decisions from $TTI=1 \quad to \quad T$

\end{algorithmic}

\end{algorithm}

\subsection{Baseline Algorithms}

In this section, two baseline algorithms are introduced that have been used for the performance comparison. The first baseline algorithm for RAT selection is based on a predefined threshold \cite{1}. This is called the heuristic baseline. Here, the threshold is calculated for each UE based on the metrics like load at eNB $(l_e)$ and gNB $(l_g)$, channel condition of a user under LTE $(ch_{e,u})$ and 5G BS $(ch_{g,u}$), service type of a user $(S_u)$. The channel condition is determined to be good or bad considering a threshold of received SINR values. Similarly, the load at each RAT is determined based on a threshold value. Based on the mentioned metrics, a value $T_u$ is calculated that is used for selecting the RAT for a UE after comparing it with a predetermined threshold $(T_{th})$. Following equation is used to calculate the value for $T_u$:
\begin{equation}
    T_u(l_e,l_g,ch_{e,u},S_u)=\alpha l_e+\beta l_g+\gamma ch_{g,u}+\delta S_u,
\end{equation}
where $\alpha$, $\beta$, $\gamma$, and $\delta$ are the weights associated with considered parameters that can be modulated based on the impact of any certain metric on system performance. $T_{th}$ is set to be the mean of all the possible values of $T_u$. The decision of steering traffic to a particular RAT is taken the following way:
\begin{equation}
  R_u=\begin{cases}
    1, T_u>T_{th}\quad \text{(1 represents gNB)}\\
    0, T_u \leqslant T_{th}\quad \text{(0 represents eNB)}.
  \end{cases}
\end{equation}

The Q-learning algorithm has been used as another baseline in this work \cite{3}. The goal is to investigate how DQN performs against the Q-learning algorithm.

\section{Performance Evaluation}
\label{s5}
\subsection{Simulation setup}
We have conducted MATLAB based simulations considering 1 eNB and 4 gNBs with 30 users in total. There are a total of 1 macro-cell and 4 small cells facilitated by the gNBs and an eNB. A macro-cell and a small-cell have carrier frequencies of 3.5 GHz and 0.8 GHz respectively. Specifications of the traffic classes used in this study have been summarized in TABLE I. For the experimental results, the load has been varied between 5-10 Mbps. Proportion of the voice, video, and gaming traffic is 20\%, 50\%, and 30\% respectively. Higher proportion of the video traffic is deliberately considered to observe how the system performs with the higher throughput requirements. Also, gaming traffic has the most stringent delay requirement and we wanted to see if the system performs well enough to meet such precise requirement. Therefore it has a higher percentage compared to the voice traffic. QoS requirements associated with delay and throughput for the three types of traffic classes are specified based on the existing literature \cite{14} and 3GPP specifications (see TABLE I.). We are using multi-RAT dual connectivity architecture, an NSA mode where LTE and 5G NR BSs serve together. An architecture specified in \cite{13} has been used where the dual connectivity is ensured via evolved packet core \cite{15}. Transmission power of the LTE BS and 5G NR BSs are set to 40W and 20W. Furthermore, bandwidth for the LTE and 5G RAT are fixed to 10MHz and 20MHz.

\begin{table}[hbt!]
\caption{Traffic Class Description and Simulation Settings}
\begin{center}
\renewcommand\arraystretch{1.2}
\begin{tabular}{|cc|}
\hline
\multicolumn{1}{|c|}{\textbf{\begin{tabular}[c]{@{}c@{}}Traffic class specification\end{tabular}}} & \textbf{Values} \\ 
\hline
\multicolumn{1}{|c|}{Traffic model}                                          & \begin{tabular}[c]{@{}c@{}}Poisson distribution, video\\  and gaming traffic {[}14{]}\end{tabular} \\ 
\hline
\multicolumn{2}{|c|}{\textbf{Voice traffic}}\\ 
\hline
\multicolumn{1}{|c|}{Packet size} & 30 bytes\\ 
\hline
\multicolumn{1}{|c|}{$T_{QoS}, D_{QoS}$} & 0.1 Mbps , 100ms \\ 
\hline
\multicolumn{1}{|c|}{Proportion of the traffic} & 20\%  \\ 
\hline
\multicolumn{2}{|c|}{\textbf{Video traffic}}\\ 
\hline
\multicolumn{1}{|c|}{Packet size} & 250 bytes \\ 
\hline
\multicolumn{1}{|c|}{$T_{QoS}, D_{QoS}$} & 10 Mbps, 80ms \\
\hline
\multicolumn{1}{|c|}{Proportion of the traffic}  & 50\% \\
\hline
\multicolumn{2}{|c|}{\textbf{Gaming traffic}} \\
\hline
\multicolumn{1}{|c|}{Packet size  (gaming traffic)} & 120 bytes\\ 
\hline
\multicolumn{1}{|c|}{$T_{QoS}, D_{QoS}$} & 5 Mbps, 40ms \\ 
\hline
\multicolumn{1}{|c|}{Proportion  of the traffic} & 30\%  \\ 
\hline
\end{tabular}
\end{center}
\vspace{-2em}
\end{table}

\subsection{Simulation results}

The performance of the proposed algorithm is evaluated in terms of two KPIs: Average system throughput and network delay. In Fig 3, we present a comparison in terms of system throughput under different user loads. The proposed DQN outperforms heuristic and Q-Learning baselines by gaining 6\% and 10\% increased throughput, respectively. 

\begin{figure}[!t]
\centerline{\includegraphics[width=0.65\linewidth]{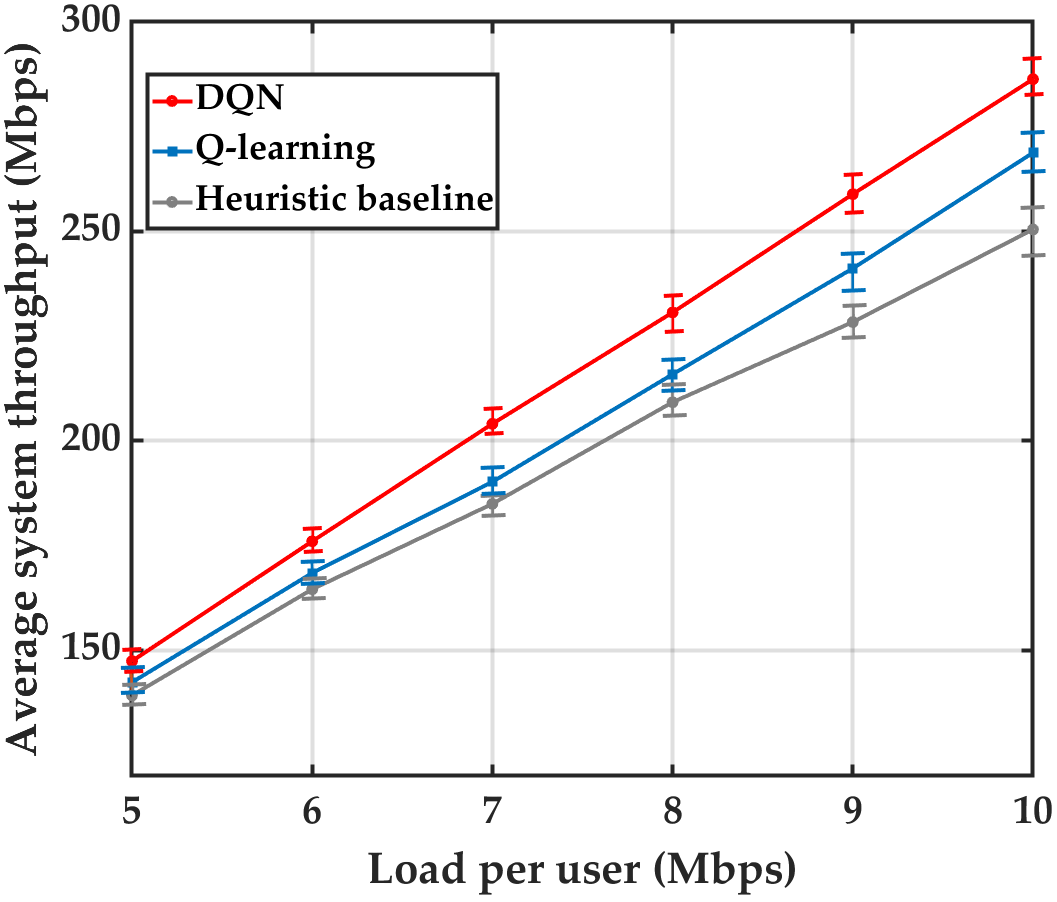}}
\caption{System throughput against traffic load.}
\label{fig}
\vspace{-1em}
\end{figure}

Fig. 4 presents the performance comparison of the proposed DQN-based traffic steering method with the other baselines in terms of delay. The DQN-based method achieves 23\% and 33\% decrease in network delay compared to the baselines. Note that, the proposed method and the Q-learning, both have a reward function formulated based on throughput and delay. Whenever high delay is experienced for steering traffic to a particular RAT, the system learns. That is why, both of them have better performance compared to the heuristic baseline. In Fig. 4, delay is calculated considering all the traffic classes together at each load. 

It should be mentioned that the main reason of the improved performance of the proposed method is the use of DQN, that outperforms Q-learning in terms of exploration efficiency and achieves higher average reward. Q-learning suffers due to longer exploration period and gets lower average reward since it does not have a DNN as an approximator which compels the agent to cover larger state and action space. 

\begin{figure}[!t]
\centerline{\includegraphics[width=0.65\linewidth]{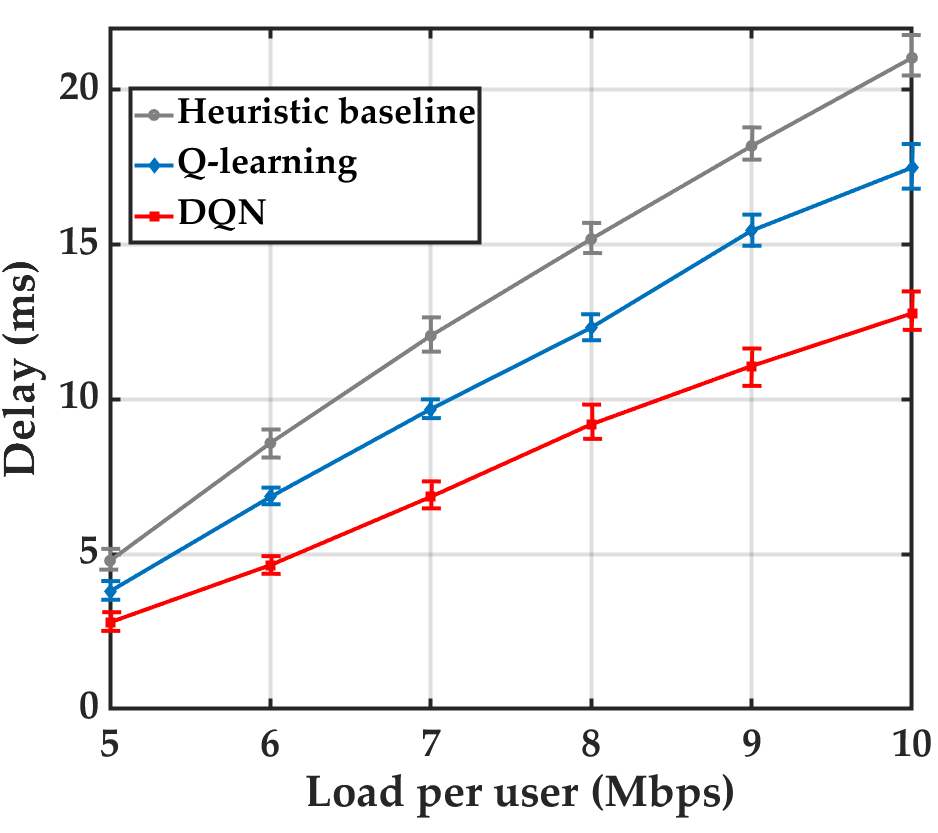}}
\caption{System delay against traffic load.}
\label{fig}
\vspace{-1.2em}
\end{figure}

\begin{figure}[!t]
\centerline{\includegraphics[width=3.4in]{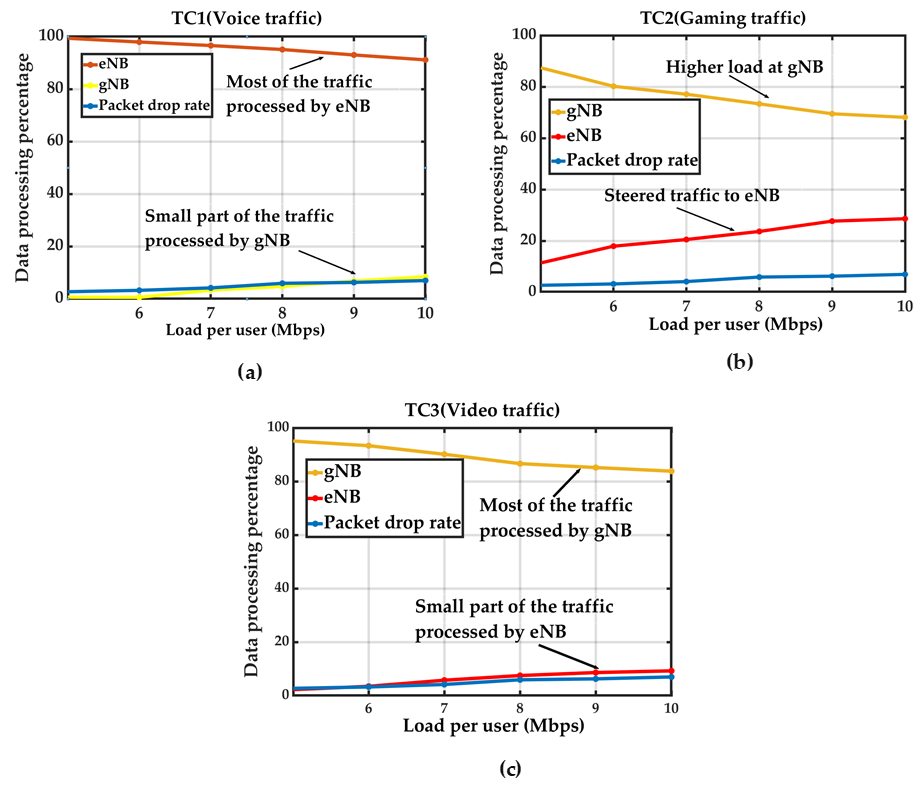}}
\caption{Data processing percentage for different traffic types.}
\label{fig}
\vspace{-1em}
\end{figure}

In this work, we also want to steer a particular type of traffic to a specific RAT. For example, steering the voice traffic constantly to a gNB is a waste of resources since the throughput requirement is not that high for such traffic. Fig. 5 is presented which shows what percentage of a traffic class is processed by a particular RAT and when the traffic gets steered due to higher load. In Fig. 5(a), it is observed that most of the voice traffic is processed by the eNB, however, a small portion of the traffic is processed by the gNB too whenever the system experiences higher load. For the video and gaming traffic, it is observed that most of the traffic is processed by the gNB.

Lastly, Fig. 6 demonstrates how traffic steering occurs whenever a high load is experienced in a BS with a particular RAT. We start with one UE at the 300th time slot and increase the number of UEs in a small cell up to six for different traffic classes. The variable $L$, in the respective figure represents load in terms of queue length. At the 1800th time slot, it can be seen that four among six UEs are steering different types of traffic to the 5G NR BS. This results in higher load and we can see that the third and fourth UEs are experiencing high load (value of $L$ changed from 0 to 1). So, in the next observed time slot, these two UEs steer the traffic to the eNB. In the 2100th time slot, we can see four UEs steering voice, video, and gaming traffic to the only eNB in our system. This incurs high load at eNB and in the next observed slot we can see that the sixth UE has switched its traffic to the gNB.

\begin{figure}[!t]
\centerline{\includegraphics[width=0.65\linewidth]{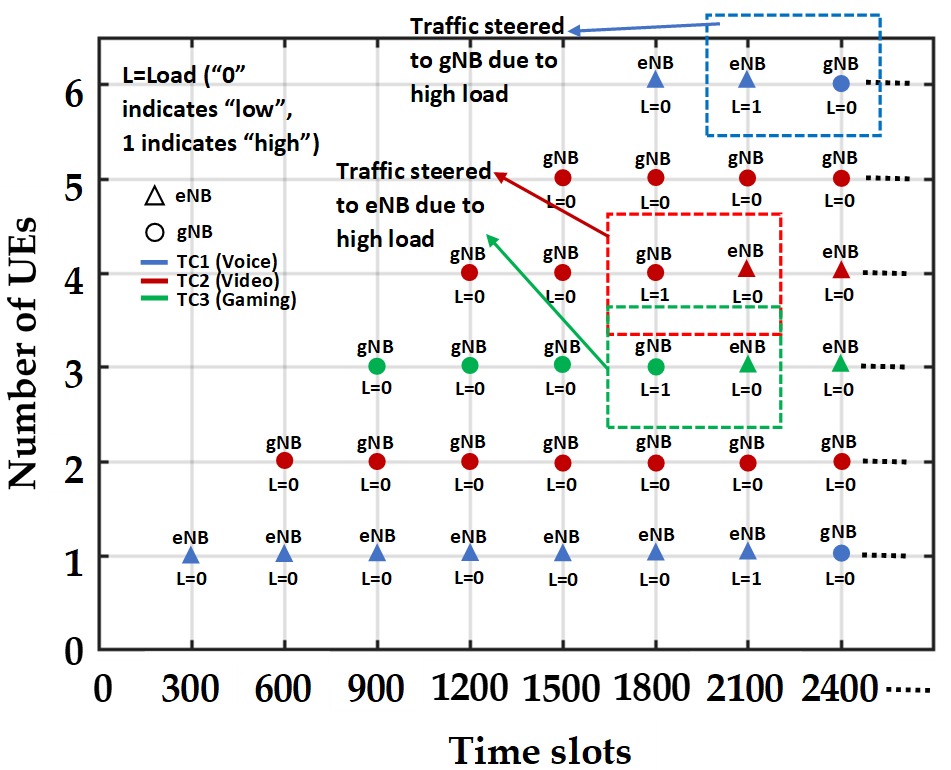}}
\caption{Traffic steered to other RAT as load changed.}
\label{fig}
\vspace{-2em}
\end{figure}

\section{Conclusions}
\label{s6}
In this study, we have proposed a novel method that can perform RAT specific and QoS aware traffic steering using DQN. It gains 6\% and 10\% increase in average system throughput compared to the Q-learning and heuristic-based baseline respectively. Moreover, it achieves 23\%
and 33\% times decrease in network delay compared to the baselines. Apart from the better performance in terms of the KPIs, the proposed method can perform RAT specific traffic steering ensuring efficient use of network resources. Lastly, the proposed DQN-based traffic steering can successfully perform load balancing in an optimal way as whenever high load is induced to a particular RAT, traffic is steered to another RAT dynamically. 

\section*{Acknowledgement}
This work has been supported by MITACS and Ericsson Canada, and NSERC Collaborative Research and Training Experience Program (CREATE) under Grant 497981.

\bibliographystyle{IEEEtran}
\bibliography{ref.bib}
\end{document}